\documentclass{vgtc}                          

\graphicspath{{figures/}{pictures/}{images/}{./}} 

\usepackage{times}                     

\usepackage{tabu}                      
\usepackage{booktabs}                  
\usepackage{lipsum}                    
\usepackage{mwe}                       

\usepackage{mathptmx}                  

\onlineid{0}

\vgtccategory{Research}

\vgtcinsertpkg

\title
{Visual Tools for Input and Reflection in Social Work}

\author
{Alexander\,Rind\thanks{e-mail: alexander.rind@fhstp.ac.at} 
\and Julia\,Boeck
}
\affiliation{St. P{\"o}lten University of Applied Sciences, St. P{\"o}lten, Austria}

\abstract{
Social workers need visual tools to collect information about their client's life situation, so that they can reflect it together and choose tailored interventions. easyNWK and easyBiograph are two visual tools for the client's social network and life history.
We  recently redesigned both tools in a participatory design project with social work faculty and professionals.
In this short paper we discuss these tools from perspective of input visualization systems.

} 


\begin{document}

\maketitle

\section{Introduction} 

Social work as a profession is dedicated to achieving social inclusion and integration of people in distressing situations.
When social work professionals (SWP) counsel a client, they need to gather information about the client's life situation, their complex problems, strengths, and resources. On this basis, they can prioritize problems and  provide tailored support that takes advantage of the client's resources. 
Thus, they apply methods of `social diagnostics' \cite{pantucek_2005_soziale} usually at the start of counseling and, depending on their specialty, repeat these diagnostics later.  
Over the last hundred years, a number of social diagnostics tools have been developed that are characterized by an interweaving of graphical representations with qualitative information collected through interviews. These tools range from simple hand drawings of relationship lines with paper and pencil or physical representations with symbols such as coins, stones, game pieces to complex computer programs. 
Yet, professionals have been looking for tools that are easy to use, tailored for social diagnostics, and support editing. 

\begin{figure*}
    \centering
    \includegraphics[width=1\linewidth]{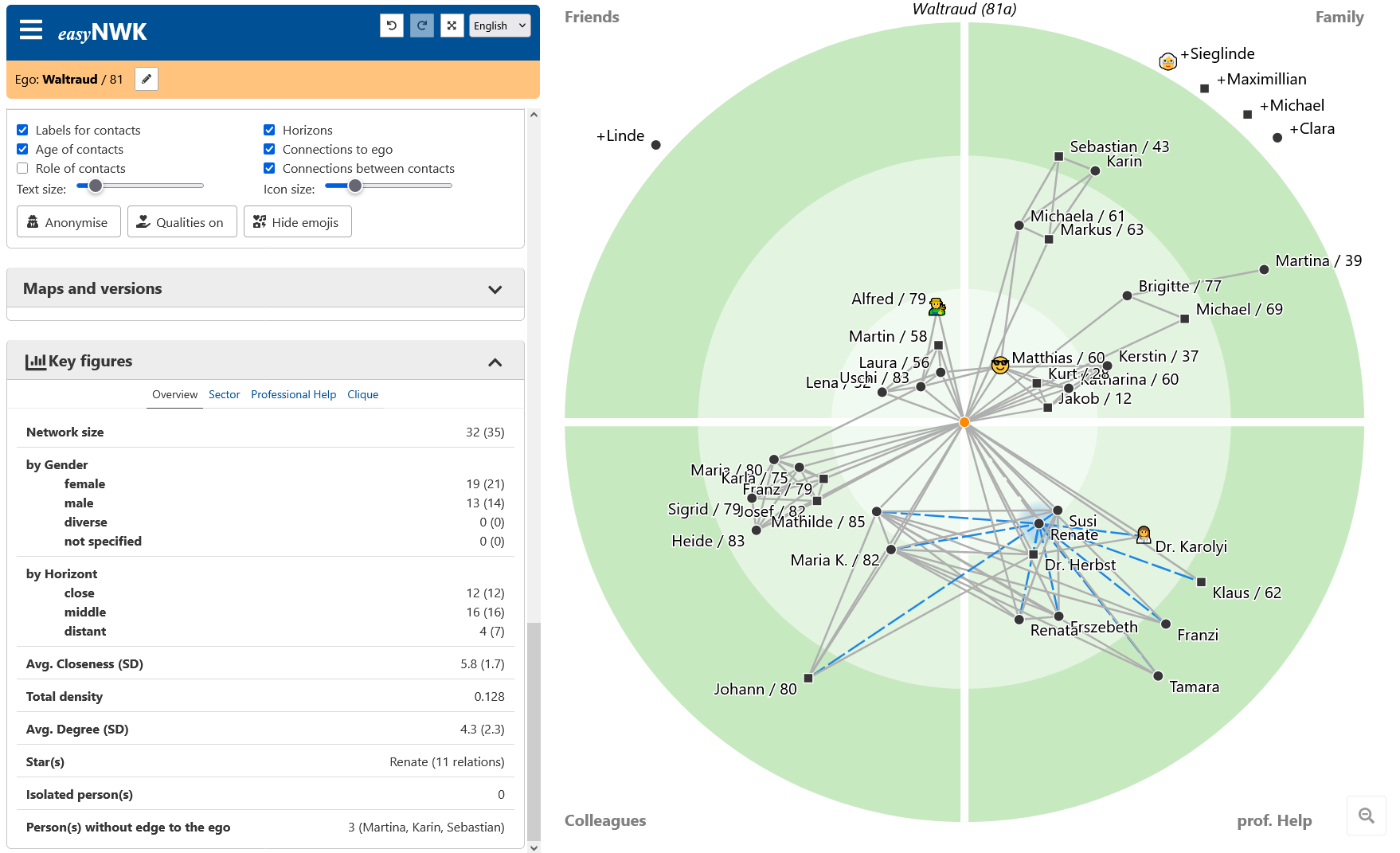}
    \caption{easyNWK with the social network of the fictitious client Waltraud (female, 81 years old).}
    \label{fig:easyNWK}
\end{figure*}

\begin{figure*}
    \centering
    \includegraphics[width=1\linewidth]{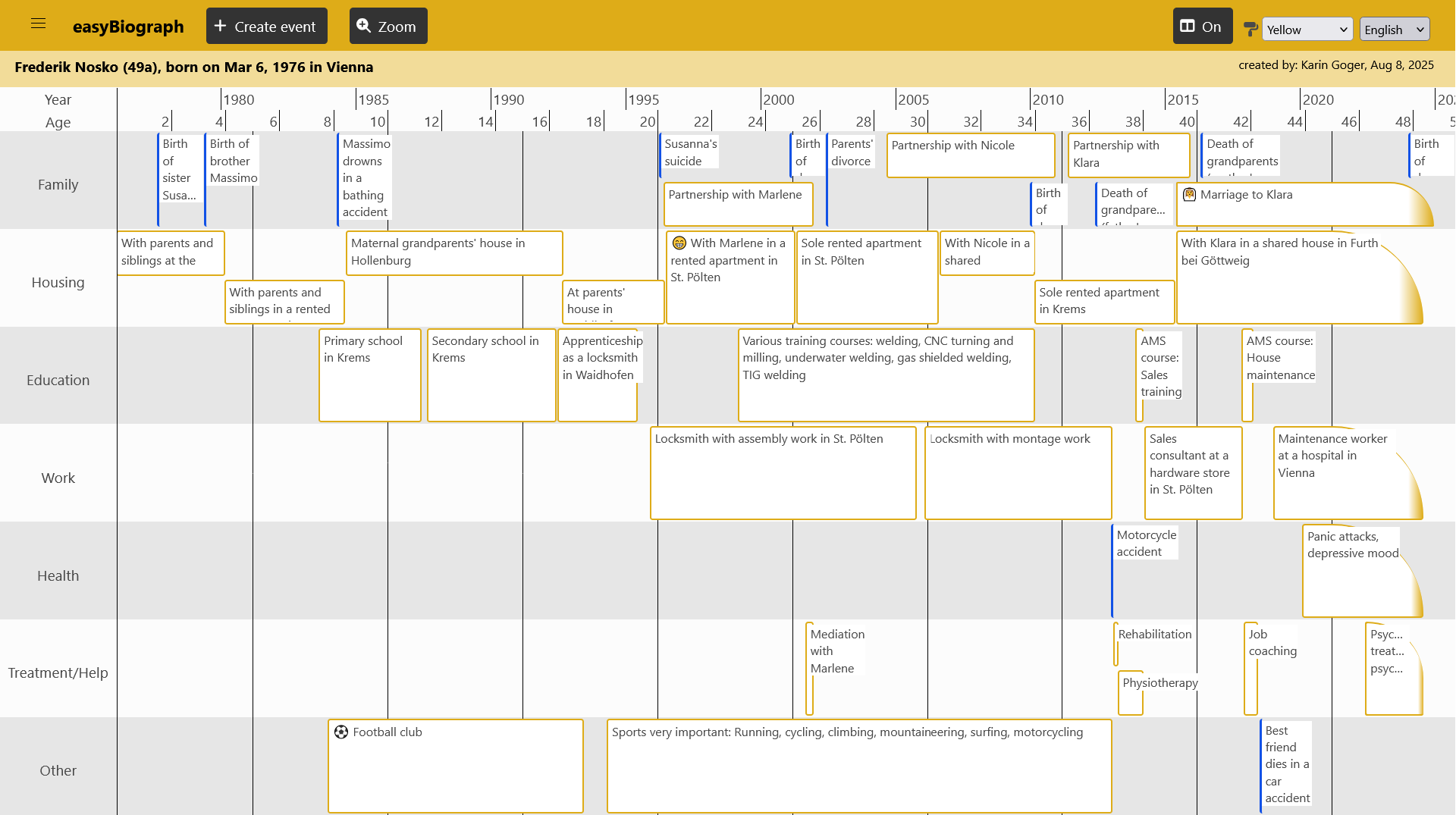}
    \caption{easyBiograph with the time bar of the fictitious client Frederik Nosko (male, 49 years old).}
    \label{fig:easyBiograph}
\end{figure*}

This work focuses on visual tools for two dimensions of the client's situation, which we worked on in a recent redesign project:
\begin{itemize}
    \item easyNWK (\cref{fig:easyNWK}) \cite{rind_2024_easynwk} for their social capital and 

    \item easyBiograph (\cref{fig:easyBiograph}) \cite{bock_2024_easybiograph} for their life history.
\end{itemize}

This short paper briefly reports on the methods and results of the redesign. After that, it reflects these tools from perspective of input visualization systems.

\section{Methods}

Both tools are based on social diagnostics methods that have been formalized by Pantu{\v c}ek-Eisenbacher \cite{pantucek_2005_soziale} and are widely used in German-speaking countries. 
Since previous Java-based tools have had issues of  usability and deployability,
we redesigned easyNWK and easyBiograph from scratch as web apps.
The redesign followed a participatory design process with core team of four professors of social work in Austria, Germany, and the Netherlands and two visualization researchers (AR, JB).
Several focus groups with other social work faculty and professionals broadened the spectrum of input and feedback. 
In addition iterative versions of the web apps were tested by cohorts of students.
Subsequently, the core team discussed and prioritized the feedback.

The redesigned easyNWK and easyBiograph were implemented as a web application using the JavaScript libraries vue.js and D3.js.

\section{Results}

The easyNWK web app implements the egocentred network map as conceptualized by Pantuček-Eisenbacher \cite{pantucek_2005_soziale}.
This concept combines three aspects from various previous approaches in sociology/social work: 
Distance from the center (= ego) is adopted from the social atom,
the relationships between contacts have been encoded as a node-link network in the sociogram,
and thematic sectors (e.g., family) have been used in various previous social network maps.
Thus, it provides a suitable standard configuration for practice in social work. 
While collecting the client’s network contacts, the SWP positions the contacts as marks in a polar coordinate system with the mark representing the client in the origin, similar to an ego-network visualization \cite{ehlers_2024_me}. In some cases, minimal data entry of a (nick-)name and a position on the network map are sufficient -- in other cases, additional metadata such as gender, role and age are desired.
The easyNWK web application  treats all metadata fields as optional except for the contacts' (nick-)names. It shows the network map on the right and a list of contacts on the left. Contacts can be added or edited with a double click on the network map or with buttons on the list.
It provides a list of key metrics suitable for the social work context, creation of multiple versions of the network map, and optionally emoji as contact symbols.

The easyBiograph web also implements a social diagnostic approach conceptualized by Pantuček-Eisenbacher \cite{pantucek_2005_soziale}.
Like LifeLines \cite{lindwarmalonso_1998_viewing}, it collects events of a client's life history along a horizontal time axis that is labeled with the calendar year as well as the client's age.
The vertical space is distributed into six or more swim lanes corresponding to standardized and freely defined categories (Family, Housing, Education, Work, Health, Treatment/Help). 
Events are added with direct manipulation  by dragging over the time interval in swim lane where they belong. 
It provides text fields for documentation and the possibility to mark events with an emoji.

\section{Reflection}

Having worked on the participatory design of easyNWK and easyBiograph allows us to reflect on some of the outcomes and draw implications for future work, especially on visualization as an input mechanism.

The SWP and the client use the web apps \textit{collaboratively} in a counseling session dedicated to  social network or their life history. Typically, they would sit together in front of a computer and the SWP would enter the data that the client mentions in their conversation.
Both can use the visual display:
clients can point on a location in the visual display to indicate where contacts or life events should be placed, and SWP can specifically ask about sparsely filled locations. 
This conversational setting has an effect on the data collected. For example, the number of contacts in a network map depends not only on the client's situation but also on how long and how skilled the SWP asks them to think about acquaintances. Such  weak-tie contacts can be important for professional social work contexts such as job search. 
Yet, this makes network metrics hardly comparable. 
Another topic for future research is how tensions between SWPs and clients  occur in these sessions and how they are resolved.

    The \textit{standardization and customizability} of the input categories have been a continuous topic of discussions. 
    On the one hand professionals expressed an intent to adapt the tools for the specific context of their clients, on the other common categories and consistent layout are needed when comparing different network maps or time bars.
    For example, a compromise could be reached in easyBiograph with the top six category swim lanes standardized and additional custom category swim lanes could be added below.
    In easyNWK, several professionals reported that it was important for their clients that they can also place pets on the network map even though only living humans can, by the underlying social work theory, contribute to social capital. This was resolved with a special category of non-human contacts that can be placed on the map but are not counted in the metrics.
    Other fields such as the contact's role allow for free text entry in addition to predefined roles.

In the discussions 
with social work experts several  \textit{unexpected design decisions} emerged and were subsequently implemented. 
For example, the first design mockups of easyBiograph featured events that all had the same consistent height and category swim lanes were of different size depending on their density. However, social work experts conceptualized all swim lanes as equally important and argued that co-occurrence should be shown as a split of resources (e.g., in the work swim lane, two parallel jobs take half the height of a full-time job).
In easyNWK, contacts are placed by the clients in a fixed position in the polar coordinate system, rather than being dynamically placed by a force-based layout algorithm based on their connections. This allows clients to directly express their view of their network and patterns to emerge from the input.

The \textit{performance of data input operations} was essential for usability -- not only for initial data but later editing was needed.
For example, the fixed positions in easyNWK could result in undesired clutter that clients would want to resolve by redistributing contacts manually. 
We are still experimenting with input actions to make such operations more efficient. Additionally, we are exploring semi-automated layouts and plan to evaluate their suitability in social work.

\section*{Supplemental Materials}
\label{sec:supplemental_materials}

easyNWK and easyBiograph are free and open-source software under the MIT license. Their source code is archived on Zenodo at 
\url{https://doi.org/10.5281/zenodo.14474950} and \url{https://doi.org/10.5281/zenodo.14389168}.

\acknowledgments{
The authors wish to thank Karin Goger, the TransSoDia consortium, and all social work professionals who provided us inputs. 
Additionally, Karin Goger has prepared the fictitious client data presented in \cref{fig:easyNWK} and \cref{fig:easyBiograph}.
This work was supported in part by the Erasmus+ project TransSoDia cofunded by the European Union.}

\bibliographystyle{abbrv-doi-narrow}
\bibliography{ref-easynwk}
\end{document}